\documentclass[twocolumn,showpacs,amssymb,10pt]{revtex4}
\usepackage{epsfig}
\usepackage{bm}
\usepackage{color}

\begin{document}
\newcommand{\pr}{^{\prime}}
\newcommand{\bfx}{\overrightarrow{x}}
\newcommand{\bfp}{\overrightarrow{p}}
\newcommand{\la}{\langle}
\newcommand{\ra}{\rangle}
\newcommand{\rp}{{p}}
\newcommand{\rpp}{{{p}^{\prime}}}
\newcommand{\rx}{{x}}
\newcommand{\vare}{\varepsilon}
\newcommand{\eps}{\varepsilon}
\newcommand{\be}{\begin{eqnarray}}
\newcommand{\ee}{\end{eqnarray}}
\newcommand{\ba}{\begin{array}}
\newcommand{\ea}{\end{array}}
\newcommand{\proc}{{negative-continuum dielectronic recombination }}
\newcommand{\balpha}{{\mbox{\boldmath$\alpha$}}}
\newcommand{\bnabla}{{\mbox{\boldmath$\nabla$}}}
\newcommand{\bfr}{{\bf r}}


%
%
\def\ket#1{ $ \left\vert  #1   \right\rangle $ }
\def\ketm#1{  \left\vert  #1   \right\rangle   }
\def\bra#1{ $ \left\langle  #1   \right\vert $ }
\def\bram#1{  \left\langle  #1   \right\vert   }
\def\spr#1#2{ $ \left\langle #1 \left\vert \right. #2 \right\rangle $ }
\def\sprm#1#2{  \left\langle #1 \left\vert \right. #2 \right\rangle   }
\def\me#1#2#3{ $ \left\langle #1 \left\vert  #2 \right\vert #3 \right\rangle $ }
\def\mem#1#2#3{  \left\langle #1 \left\vert  #2 \right\vert #3 \right\rangle   }
\def\rmem#1#2#3{  \left\langle #1 \left\vert \left\vert  #2
                  \right\vert \right\vert #3 \right\rangle   }
\def\threej#1#2#3#4#5#6{ $ \left( \matrix{ #1 & #2 & #3  \cr
                                           #4 & #5 & #6  } \right) $ }
\def\threejm#1#2#3#4#5#6{  \left( \matrix{ #1 & #2 & #3  \cr
                                           #4 & #5 & #6  } \right)   }
\def\sixj#1#2#3#4#5#6{ $ \left\{ \matrix{ #1 & #2 & #3  \cr
                                          #4 & #5 & #6  } \right\} $ }
\def\sixjm#1#2#3#4#5#6{  \left\{ \begin{array}{ccc}
                                               #1 & #2 & #3  \\
                                               #4 & #5 & #6
                     \end{array} \right\}   }
\def\ninejm#1#2#3#4#5#6#7#8#9{  \left\{ \matrix{ #1 & #2 & #3  \cr
                                                 #4 & #5 & #6  \cr
                         #7 & #8 & #9 } \right\}   }
\def\twobytwo#1#2#3#4{  \left| \begin{array}{cc}
                                   #1 & #2   \\[0.2cm]
                                   #3 & #4   \end{array} \right|   }

\def\mpl#1{\uppercase{#1}}                               
\def\etal{et al.}                                        %

\def\sfc#1{ \textbf{... #1 ...} }                        
\def\sfq#1{ \textbf{(?? #1 ??)} }                        
\def\sfi#1{ $\bullet\;$ \textit{#1}}                     
\def\sfl#1{ {\large $\to\;$} \textsf{#1}}                

\def\asc#1{ {\color{black} \textbf{... #1 ...} }}         
\def\asi#1{ {\color{black} $\bullet\;$ \textit{#1} }}     
\def\asr#1{ {\color{black} \textbf{I would remove this: ... #1 ...} }}  

\title{Negative-continuum dielectronic
recombination\\[0.2cm] into excited states of highly--charged ions}
\author{A.~N.~Artemyev$^{1,2}$\footnote{Corresponding author.
Email: artemyev@physi.uni-heidelberg.de}, V.~M.~Shabaev$^3$, Th.
St\"{o}hlker$^{1,2}$, and A.~S.~Surzhykov$^{1,2}$}

\affiliation{
$^1$Physikalisches Institut, Universit\"{a}t Heidelberg,
Philosophenweg 12, D--69120 Heidelberg, Germany\\
$^2$GSI Helmholtzzentrum f\"ur Schwerionenforschung GmbH,
Planckstrasse 1, D--64291 Darmstadt, Germany\\
$^3$Department of Physics, St. Petersburg State University,
Oulianovskaya 1, Petrodvorets, St. Petersburg 198504, Russia }

\date{\today}

\begin{abstract}
The recombination of a free electron into a bound state of bare,
heavy nucleus under simultaneous production of
bound--electron---free--positron pair is studied within the
framework of relativistic first--order perturbation theory. This
process, denoted as ``negative--continuum dielectronic
recombination'' leads to a formation of not only the ground but also
the singly-- and doubly--excited states of the residual helium--like
ion. The contributions from such an excited--state capture to the
total as well as angle--differential cross-sections are studied in
detail. Calculations are performed for the recombination of
(initially) bare uranium U$^{92+}$ ions and for a wide range of
collision energies. From these calculations, we find almost 75 \%
enhancement of the total recombination probability if the excited
ionic states are taken into account.
\end{abstract}

\pacs{31.30.Jv, 34.80.Lx} \maketitle

%
%
\section{Introduction}

Owing to the recent advances in heavy--ion accelerators and ion
storage rings, the electron--positron pair creation in ion--atom (or
ion--ion) collisions attracts much of today's interest both in
experiment and theory \cite{eichler:95:book,Eic05,BaH07}. The investigation of
this process is of great importance not only for a better
understanding of the physics of extremely strong electromagnetic
fields but also for the development of novel collider facilities
\cite{BaR96,BrJ07}. A large number of studies are focused,
therefore, on the \textit{bound--free} pair production in which
electron is created in an atomic (ionic) shell under a simultaneous
emission of a free positron. Experimentally, such a pair
production is usually observed in relativistic
collisions of highly--charged ions with medium-- and high--$Z$
atomic targets \cite{BeG93,VaD97,BeC98}. Special attention in these
experimental studies is paid to the \textit{total} (pair creation)
cross-sections and to their dependence on both the collision energy
and the nuclear charge of both the projectile and the target. Theoretical
analysis of the total cross-sections is not a simple task since it
usually requires the solution of the \textit{two--center} time--dependent
Dirac equation. Even though a number of approaches, such as the
coupled--channel methods \cite{reinhardt:78, muellernehler:94,
  eichler:95:book, Eic05} and the lattice
numerical solutions \cite{IoB99,BuG04}, have been developed to deal
with the two--center Dirac problem, they result in very demanding
and time--consuming calculations.

\medskip

Apart from the energetic collisions of high--$Z$ projectiles with
heavy atomic targets, an alternative and very promising route has
been offered recently to investigate the
bound--electron---free--positron production. In this process a
\textit{free} incoming electron is captured into a bound state of
heavy, highly--charged ion and the released energy is converted into
an electron--positron pair \cite{artemyev:03:pra:ncdr}. Due to the
schematic similarity to the usual dielectronic recombination of
few--electron ions, this process has been denoted as
\textit{negative continuum dielectronic recombination} (NCDR). It is
important to outline, however, that in contrast to dielectronic
recombination, the electron to be ``excited'' in NCDR is not
initially bound to an ion but an electron from the negative
continuum. Therefore, negative continuum dielectronic recombination
of a free electron with an initially \textit{bare} nucleus of charge
$Z$ results in the production of a helium--like ion and a free outgoing
positron:
\begin{equation}
   \label{NCDR_definition}
   X^{Z+}+e^{-} \, \rightarrow \, X^{(Z-2)+} +e ^{+} \, .
\end{equation}
Of course, this reaction has a threshold and becomes possible only
if the energy of incoming electron in the center of mass rest frame is
larger than the sum of the positron rest energy and the energy of
the residual helium--like ion.

\medskip

Unlike the other mechanisms of pair creation, the NCDR involves only
one--center Dirac problem which significantly simplifies its
theoretical analysis. In Ref.~\cite{artemyev:03:pra:ncdr}, for
example, the first--order perturbation approach has been developed
to study the properties of the emitted positron and the total
recombination probabilities. Within this approach, based on
relativistic Dirac's theory, detailed calculations have been carried
out for the cross-sections of the NCDR into the ground $(1s)^2$ state
of (initially) bare led Pb$^{82+}$ and uranium U$^{92+}$ ions. These
calculations have indicated that the ground--state recombination
cross-sections do not exceed the value of 30 $\mu$bar. This
makes, experimental study of the NCDR process rather difficult.
It was argued, however, that a significant enhancement
of the NCDR probability can be expected due to electron transfer
into \textit{excited} states of finally helium--like ions. Together
with the unique signature of ionic charge change by two units
accompanied by a simultaneous positron emission [cf.
Eq.~(\ref{NCDR_definition})], such an enhancement shall render the
NCDR process observable with the advent of the new generation of
heavy--ion storage rings with high beam intensities such as, for
example, at the future Facility for Antiproton and Ion Research
(FAIR) at Darmstadt \cite{GSI01}.

\medskip

Despite its impact on the planned FAIR experiments, the influence of
the excited--state recombination on the total NCDR probability has
not been studied in detail until now. A first step towards such a
study was performed by us in Ref.~\cite{artemyev:05:nimb} where the
electron transfer into some low--lying bound states of high--$Z$,
bare ions has been investigated. Relativistic calculations performed
in Ref.~\cite{artemyev:05:nimb} have confirmed the significant role
of excited--state NCDR but have faced us to the necessity of a
more \textit{systematic} analysis.

\medskip

In this contribution, we apply here the first--order perturbation
theory based on the relativistic Dirac's equation to explore a
negative continuum dielectronic recombination of a free electron
into singly-- and doubly--excited states of (initially) bare,
highly--charged ions. Special emphasis in our present study is
placed on the contributions which come from the excited--state
capture to both, the total and the angle--differential NCDR cross
sections. The basic expressions for these cross-sections, as
obtained within the framework of the independent particle model
(IPM) for the description of the (final) two--electron states, are
discussed in Section \ref{section_theory}. In particular, it is
argued that any analysis of the NCDR properties can be traced back
to the transition amplitudes which describe the interelectronic
interaction in the capture process. In
Section~\ref{sec_computation}, the evaluation of these matrix
elements is briefly outlined. Results of our fully relativistic
calculations are presented then in Section \ref{sec_results} for the
initially bare uranium ions U$^{92+}$ and for a wide range of
collision energies. The calculations have been carried out for the
NCDR into all singly--excited $(1snl_j)_J$ with $n \le 4$ and
doubly--excited $(2s2l_j)_J$ states. In order to take into account
even higher--lying states, an extrapolation of our results has been
performed to larger principal quantum numbers $n$. By making use of
such an extrapolation procedure we have found that the recombination
into excited ionic states may enhance the total cross-section by
about 75 \% when compared to the ground--state NCDR. A brief summary
of this important finding and its implication for future experiments
is finally given in Section \ref{sec_summary}.

\medskip

Relativistic units ($\hbar=m_e=c=1$) are used throughout the paper.
In the following, moreover, electron energies are always defined as
including the electron rest mass.

%
%
%
%

%
%
\section{Theory}
\label{section_theory}

\begin{figure}
\epsfig{file=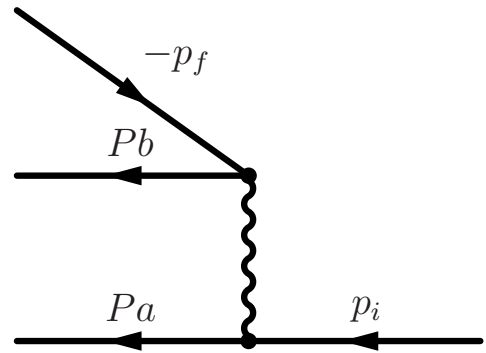,width=0.35\textwidth}
\caption{\label{schemeNCDR} Feynman diagram for the
negative-continuum dielectronic recombination. $a$ and $b$ are the
bound one-electron states and $P$ is the permutation operator. }
\end{figure}

Not much has to be said about the basic formalism for studying the
negative continuum dielectronic recombination of high--$Z$ bare
ions. Recently, this formalism, based on the quantum
electrodynamical approach, has been discussed by us in
Refs.~\cite{artemyev:03:pra:ncdr,artemyev:05:nimb}. In particular,
we argued that to zeroth order of perturbation theory, NCDR is given
by the diagram shown in Fig.~\ref{schemeNCDR}. In this diagram,
${p}_i$ is the asymptotic momentum
of the incoming electron and $a$ and $b$ denote the
quantum states of the (finally) bound electrons. Moreover, according
to the standard procedure \cite{bjorken:65,itzykson:80}, the
outgoing positron with four-momentum $p_f$ is described as
an incoming electron with four-momentum
$-p_f$.

\medskip

Evaluation of the diagram depicted in Fig.~\ref{schemeNCDR} requires
the knowledge of the (two--electron) wavefunction of the final ionic
state. Since our analysis of the NCDR process is restricted to heavy
ions for which the electron--electron interaction effects are
usually small, it is convenient to describe this wavefunction within
the framework of the independent particle model (IPM). That is, the
wavefunction of the final helium--like ion with the well--defined
angular momentum $J$ and its projection $M$ is approximated by means
of Slater determinants built from hydrogenic orbitals:
\be
   \label{wave_function_final}
\Psi^{bound}_{JM}(\bfr_1,\bfr_2)&=&N\sum_{m_a,m_b}
C_{j_am_a,\,j_bm_b}^{JM} \nonumber \\ &&\times
\left|\begin{array}{ll}
\psi_{n_a\kappa_am_a}(\bfr_1)&\psi_{n_b\kappa_bm_b}(\bfr_1)\\
\psi_{n_a\kappa_am_a}(\bfr_2)&\psi_{n_b\kappa_bm_b}(\bfr_2)\end{array}\right|
\nonumber \\ &=&
N\sum_{m_a,m_b}C_{j_am_a,\,j_bm_b}^{JM}\sum_P(-1)^P\nonumber \\
&\times &
\psi_{n_{Pa}\kappa_{Pa}m_{Pa}}(\bfr_1)\psi_{n_{Pb}\kappa_{Pb}m_{Pb}}(\bfr_2)
\,. \ee
Here, $\psi_{n_{a} \kappa_{a} m_{a}}(\bm{r})$ and $\psi_{n_{b}
\kappa_{b} m_{b}}(\bm{r})$ are the well--known solutions of the
Dirac's equation for a bound electron, $C_{j_a m_a,\,j_b m_b}^{JM}$
is the Clebsch--Gordan coefficient, $P$ is the permutation
operator, $(-1)^P$ is the permutation parity and $N$ is the
normalization factor. As usual, this factor is $N=1/2$ for $n_a =
n_b$, $\kappa_a=\kappa_b$ and $N=1/\sqrt{2}$ otherwise.

\medskip

By making use of the final state wavefunction
(\ref{wave_function_final}) we are able to evaluate the Feynman
diagram from Fig.~\ref{schemeNCDR} and, hence, to find the
differential (in positron emission angle) NCDR cross--section in the
center of mass rest--frame:
\begin{eqnarray}
\label{dcs-1}
\frac{d\sigma}{d\Omega_f}(n_a \kappa_a, n_b \kappa_b;
J) &=&
\frac{16\pi^4N^2}{v_i}{\bf p}_f^2 \nonumber \\
&& \hspace*{-3.5cm} \times \sum_{M}\sum_{m_a,\,m_b}\sum_{m_i,\,
m_f}\Big|\sum_{P}(-1)^P C^{J\,M}_{j_am_a,\,j_bm_b}\nonumber
\\ & & \hspace*{-3.5cm }\times
\mem{PaPb}{I(\vare_i-\vare_{Pa})}{{p}_i,m_i, -{p}_f,m_f}\Big|^2\, ,
\nonumber \\
\end{eqnarray}
where we assume that the incident electron is unpolarized and the spin
states of the residual ion and emitted positron remain unobserved.
In Eq. (\ref{dcs-1}), moreover, for the sake of shortness we denote $|ab
\rangle \equiv |n_a \kappa_a m_a, n_b \kappa_b
m_b\rangle$, $\vare_{i}$ and $\vare_{a,b}$ are
the one--electron Dirac energies of the incoming and bound
electrons, respectively, and $I(\omega)$ is the electron--electron
interaction operator. In Coulomb and Feynman gauges this operator
reads:
\begin{eqnarray}
\label{interelectronic}
I_C(\omega)&=&\alpha\left(\frac
{1}{r_{12}}-(\balpha_1 \cdot \balpha_2)
\frac{\exp(i|\omega|r_{12})}{r_{12}}\right. \nonumber
\\ &+&\left. \Big[(\balpha_1 \cdot
  \bnabla_1),\Big[(\balpha_2 \cdot \bnabla_2),
\frac{\exp(i|\omega|r_{12})-1}{\omega^2r_{12}}\Big]\Big]\right)\,,\nonumber
\\ \\
I_F(\omega)&=&\alpha(1-\balpha_1 \cdot
\balpha_2)
\frac{\exp(i|\omega|r_{12})}{r_{12}}\,,
\end{eqnarray}
where $r_{12}=|\bfr_1-\bfr_2|$ and ${\bm \alpha}_k$ denotes the
vector of Dirac matrices for the $k$th electron. Below, we employ
the Feynman gauge to calculate the NCDR cross-sections. It is
obvious, however, that the results obtained are gauge invariant since
both the free-- and bound--state wavefunctions in Eq.~(\ref{dcs-1})
are exact solutions of the Dirac equation.

\medskip

The transition matrix element in the last line of Eq.~(\ref{dcs-1})
contains the wavefunctions of incoming electron $\ketm{{p}_i
m_i}$ and outgoing positron $\ketm{-{p}_f m_f}$ with the well
defined asymptotic momenta. For the further evaluation of the NCDR
differential cross-section, it is therefore necessary to decompose
these continuum waves into partial waves in order to apply the
standard techniques for the theory of angular momentum. As discussed
previously, however, special care has to be taken about the choice
of the quantization axis since this directly influences the
particular form of the partial wave decomposition. Using, for
example, the incoming electron momentum ${p}_i$ as the
quantization axis ($z$--axis), the full expansion of the continuum
electron wave function is given by \cite{eichler:95:book,rose:61}:
\begin{eqnarray}
\label{electron}
|{p}_i
m_i\ra&=&\frac{1}{\sqrt{4\pi}}\frac{1} {\sqrt{\vare_i|{\bf
p}_i|}}\sum_\kappa i^l\exp(i\Delta_\kappa) \nonumber \\
&&\times\sqrt{2l+1} C^{j\,m_i}_{l\,0,\,(1/2)\,m_i} \ketm{\vare_i,
\kappa, m_i}  \,,
\end{eqnarray}
where the summation runs over Dirac's angular momentum quantum
number $\kappa = \pm (j + 1/2)$ for $l = j \pm 1/2$ with $l$
representing the parity of the partial waves $\ketm{\vare_i, \kappa,
m_i}$. In expansion (\ref{electron}), moreover, $\Delta_\kappa$ is
the Coulomb phase shift and the partial waves
\begin{eqnarray}
  \label{radial_angular_representation}
  \langle {\bf r} |\vare_i, \kappa, m_i\rangle= \left( \ba{c}
  g_{\vare_i,\kappa}(r)\Omega_{\kappa m_i}(\hat{\bf r})\\
  if_{\vare_i,\kappa}(r)\Omega_{-\kappa m_i}(\hat{\bf r})
  \ea
  \right)\, ,
\end{eqnarray}
decompose in a standard way into a radial and an angular part and,
thus, help to carry out the integration over all angles in the NCDR
cross-section (\ref{dcs-1}) analytically.

\medskip

In contrast to the incoming electron, the outgoing positron does not
necessarily move along the $z$--axis. In this case it is rather
convenient to retain the axis of spin quantization in the direction
of $z$--axis and just rotate the space part of the wavefunction
\cite{eichler:95:book}:
\begin{eqnarray}
\label{positron} |-{p}_f,m_f\ra&=&\frac{1}{\sqrt{\vare_f|{\bf
p}_f|}}\sum_{\kappa,M_f} i^{l^\prime}\exp(i\Delta_\kappa)\nonumber
\\ &&\times C^{j\,M_f}_{ l^\prime\,m_{l^\prime},\,(1/2)\,m_f}
Y^*_{l^\prime m_{l^\prime}}(-\hat{p}_f)\nonumber \\ &&\times
|-\vare_f, \kappa, M_f\rangle\, .
\end{eqnarray}
Indeed, the value of $m_f$ in this expansion does not have any
physical meaning since for the relativistic continuum electron
(positron) the spin projection has a sharp value only in the
direction of propagation (so called helicity). However, since in
Eq.~(\ref{dcs-1}) we perform summation over all possible spin states
of the outgoing positron the use of expansion (\ref{positron}) is
well justified.

\medskip

By inserting now partial wave expansions (\ref{electron}) and
(\ref{positron}) into Eq.~(\ref{dcs-1}), we finally can write the
differential NCDR cross-section as:
\begin{eqnarray}
   \label{dcs-2}
   \frac{d\sigma}{d\Omega_f}(n_a\kappa_a,n_b\kappa_b;J)
   &=&\frac{4\pi^3|{\bf p}_f|N^2} {\vare_f{\bf p}^2_i}\nonumber \\
   & & \hspace*{-2.5cm} \times \sum_{M}\sum_{m_a m_b} \sum_{m_i,
   m_f}\Big|\sum_{P,\kappa_i,\kappa_f,M_f} (-1)^P \nonumber \\
   && \hspace*{-2.5cm} \times
   C^{J\,M}_{j_am_a,\,j_bm_b}
   \;i^{l_i+l_f^\prime}\exp(i\Delta_{\kappa_i}+i\Delta_{\kappa_f})
   \sqrt{2l_i+1}
   \nonumber \\
   && \hspace*{-2.5cm} \times
   C^{j_i\,m_i}_{l_i\,0,\,(1/2)\,m_i}
   C^{j_f\,M_f}_{l_f^\prime\,m_{l_f^\prime},\,(1/2)\,m_f}
   Y_{l_f^\prime ,m_{l_f^\prime}}^*(-\hat{\bf p}_f)
   \nonumber \\
   && \hspace*{-2.5cm} \times
   \la PaPb|I(\vare_i-\vare_{Pa})| (\vare_i,\kappa_i,m_i)
   (-\vare_f,\kappa_f,M_f)\ra
   \Big|^2\,. \nonumber \\
\end{eqnarray}
Within the framework of independent particle model this equation is
still exact and allows to calculate the cross-section for the
negative continuum dielectronic recombination into any possible
final configuration of the helium--like ion. In the next Section, for
example, we make use of Eq.~(\ref{dcs-2}) to investigate electron
capture into singly-- as well as doubly--excited states of U$^{90+}$
ions.

\begin{figure}
\epsfig{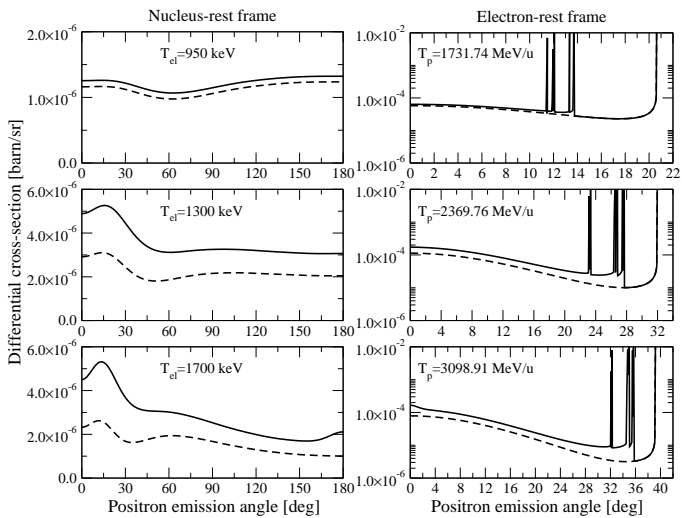}
\caption{\label{Fig2} Differential cross-section in the emitted
positron angle for the NCDR into $(1s)^2$ state (dashed line) and
sum of the cross-sections into all possible $(1s nl_j)$ with $n\le
4$ and $(2s 2l_j)$ states (solid line).}
\end{figure}
\section{Computation}
\label{sec_computation}

As seen from Eq.~(\ref{dcs-2}), any further analysis of the NCDR
angle--differential cross-sections can be traced back to the
transition amplitude:
\begin{eqnarray}
   \label{transition_amplitude}
   M(a, b) &&\nonumber \\
   && \hspace*{-1.5cm} = \mem{n_a \kappa_a m_a, n_b \kappa_b
   m_b}{I(\vare_i-\vare_{a})}{(\vare_i,\kappa_i,m_i)
   (-\vare_f,\kappa_f,M_f)} \nonumber \\
\end{eqnarray}
which describes the interelectronic interaction during the capture
process. In the computations below, we make use of the
single--particle Dirac's wavefunctions modified for the
\textit{finite} nuclear size in order to evaluate such an amplitude.
Similar to our previous work \cite{artemyev:03:pra:ncdr}, the radial
components of these wavefunctions have been calculated with the help
of the RADIAL package \cite{salvat:95:cpc}. This package can be
applied for computation of both bound and continuum solutions of the
Dirac equation, but only for positive energies. In order to find
negative--continuum solutions, we used the fact that the large
$g_{-\eps,-\kappa}^{(e^-)}$ and the small
$f_{-\eps,-\kappa}^{(e^-)}$ radial components of the electron wave
function can be expressed in terms of the positron function
components as
$g_{-\eps,-\kappa}^{(e^-)}(r)=f_{\eps,\kappa}^{(e^+)}(r)$,
$f_{-\eps,-\kappa}^{(e^-)}(r)=g_{\eps,\kappa}^{(e^+)}(r)$
\cite{rose:61}.

\medskip

By making use of the standard two--component representation
(\ref{radial_angular_representation}) of the electron (and positron)
wavefunctions as well as the multipole expansion of the
interelectronic interaction operator (\ref{interelectronic}), we
were able to separate NCDR transition amplitude
(\ref{transition_amplitude}) into the radial and the spin--orbital
parts. While the latter can be evaluated analytically using the
calculus of the irreducible tensor operators \cite{balashov:00}, the radial
integrations have been accomplished numerically with the help of
Gauss--Legendre quadratures. To perform this
two-dimensional integration we used the standard technique, elaborated for the
calculation of such integrals in the case when all four wave function describe
bound electrons (see e.g. Ref. \cite{johnson:88:mbpt}). However, for
relativistic ion--atom collisions such an integration is not a simple task due
to the rapid oscillations of the Dirac continuum wavefunctions. In
order to overcome this problem we had to use much denser grid of
integration knots, than in the case of bound wave functions. Most
seriously these oscillations affect NCDR calculations for the
highly--excited ionic states which have a significant overlap with
the continuum spectrum. In order to achieve a required accuracy of
0.01 \% in the computation of radial integrals for these states, we
have used few hundreds integration grid points in the
Gauss--Legendre procedure. Stability of our numerical results with
respect to the exact number and position of radial knots has been
proved in series of calculations.

Apart from the influence of the particular integration
(Gauss--Legendre) scheme, the role of nuclear--size effect in the
numerical evaluation of the NCDR cross sections has been also
verified. By making use of the RADIAL package, calculations have
been performed not only for the (finite--size) nucleus with
root-mean-square radius of 5.86 fm but also for the point--like nucleus.
Nuclear--size effect, found from this calculations does not exceed
the level of 0.1\% which is well below experimental accuracy of the
planned NCDR experiments.

\medskip

\section{Results and discussion}
\label{sec_results}

After a brief discussion of the theoretical background and the main
computational procedures we are ready now to focus on the question
of how the electron capture into (highly) excited ionic states
affects the \textit{total} NCDR cross-section $\sigma_{tot}$. We
define such a cross-section as a sum over all partial cross-sections
describing recombination into possible particular states
of the (finally) helium--like projectile:
\begin{eqnarray}
   \label{cross_section_total}
   \sigma_{tot} &=& \sum\limits_{n_a \kappa_a, n_b \kappa_b; J}
   \sigma(n_a \kappa_a, n_b \kappa_b; J) \, ,
\end{eqnarray}
where, in turn, $\sigma(n_a \kappa_a, n_b \kappa_b, J)$ is obtained
upon integration of Eq.~(\ref{dcs-2}) over the positron emission angle:
\begin{eqnarray}
   \label{cross_section_partial}
   \sigma(n_a \kappa_a, n_b \kappa_b, J) =
   \int{\frac{d\sigma}{d\Omega_f}(n_a \kappa_a, n_b \kappa_b;
   J) \, {\rm d}\Omega_f} \, .
\end{eqnarray}
Most naturally, this angular integration can be performed
analytically by using the explicit form of the differential cross
section (\ref{dcs-2}) and the orthogonality properties of the
spherical harmonics.

Before presenting results of our calculations for the total NCDR
cross sections, first we would like to discuss the influence of the
excited--state recombination on the emission pattern of positrons.
In Fig.~\ref{Fig2}, we display the differential cross sections
(\ref{dcs-2}) as a function of positron emission angle for the
recombination into helium--like U$^{90+}$ ions. In the left column
of the figure results of our fully relativistic calculations are
presented in the rest frame of the ion which---to zeroth order in
$m_{el}/M_{ion}$---coincides with the center of mass rest--frame.
Within this frame we have studied positron emission patterns for
three kinetic energies of incoming electron: 950 keV (upper panel),
1300 keV (middle panel) and 1700 keV (lower panel). For all three
energies we present not only the $K$--shell capture cross section
(dashed line) but also the sum of the partial (differential)
cross-sections for the NCDR into all singly excited states $(1s \,
nl_j)_J$ with $n \le 4$ and doubly--excited states $(2s \, 2l_j)_J$.
As seen from Fig.~\ref{Fig2}, while the capture into these states
almost does not affect the NCDR cross-section at the near--threshold
energy of $T_{el}$ = 950 keV, it results in a significant
enhancement of positron emission for higher electron velocities. For
the collision energy of $T_{el} = 1700$ keV, for example, the NCDR
cross-section increases by about factor of two if the recombination
into excited ionic states is taken into account.

\medskip

Until now we have discussed the NCDR differential cross-sections
only in the rest frame of the ion. While the theoretical analysis of the
electron recombination process is easier to be performed in such a
\textit{projectile} frame, the forthcoming experimental results on
the positron emission most likely will be obtained in the electron
rest--frame (\textit{laboratory} frame). In the following,
therefore, we shall apply the Lorenz transformation in order to find
the cross-section (\ref{dcs-2}) \textit{in the laboratory frame}. As
it was already discussed in
Refs.~\cite{artemyev:03:pra:ncdr,eichler:95:book,dedrick:62}, such a
transformation predicts that positron emission in the laboratory
frame is possible only within the forward hemisphere and that the
maximal emission angle is given by:
\begin{eqnarray}
   \label{maximal_angle}
   \sin(\theta_{\rm max})=\frac{\beta_{p}\gamma_{p}}{\beta_{n}\gamma_{n}}\,
   .
\end{eqnarray}
Here, $\gamma=1/\sqrt{1-\beta^2}$ is the relativistic Lorenz factor
and $\beta$ denotes the velocity (in units of the speed of light).
Moreover, the values with sub--index $''p''$ correspond to the
positron's speed and Lorenz factor in the nucleus--rest frame while
ones with sub--index $''n''$ to the nucleus in the laboratory frame.

\begin{table}
\caption{\label{tabU}  The total cross-section of NCDR
 into the $(1s nl_j)$, and $(2s2l_j)$
states of U$^{90+}$  (initial charge state 92+). Units are keV for
the  kinetic energy of the electron, $\mu$barn for the cross
sections.}
\begin{ruledtabular}
\begin{tabular}{|l|l|l|l|}
\hline $T_{el}$&950&1300&1700 \\ \hline
$(1s)^2$& 13.8& 26.7& 20.0\\ 
$(1s2s)$& 6.82$\times 10^{-1}$& 7.47                & 5.91\\ 
$(1s2p)$& 2.62$\times 10^{-1}$& 2.87                & 2.41\\ 
$(1s3s)$& 9.49$\times 10^{-2}$& 2.12                & 1.73\\ 
$(1s3p)$& 4.02$\times 10^{-2}$& 9.58$\times 10^{-1}$& 8.18$\times 10^{-1}$\\ 
$(1s3d)$& 7.80$\times 10^{-4}$& 1.37$\times 10^{-2}$& 1.88$\times 10^{-2}$\\ 
$(1s4s)$& 2.38$\times 10^{-2}$& 8.40$\times 10^{-1}$& 6.95$\times 10^{-1}$\\ 
$(1s4p)$& 1.22$\times 10^{-2}$& 4.16$\times 10^{-1}$& 3.54$\times 10^{-1}$\\ 
$(1s4d)$& 3.33$\times 10^{-4}$& 8.89$\times 10^{-3}$& 7.41$\times 10^{-3}$\\ 
$(1s4f)$& 2.61$\times 10^{-5}$& 2.97$\times 10^{-4}$& 1.01$\times 10^{-3}$\\ 
$(2s)^2$&                     & 4.69$\times 10^{-1}$& 4.41$\times 10^{-1}$\\
$(2s2p)$&                     & 3.38$\times 10^{-1}$& 3.37$\times
10^{-1}$\\ \hline $\Sigma$   & 14.9                & 42.2 & 32.7\\
\hline Extrapolated         & 14.9   & 44.3(7) & 34.5(9)\\
\hline
\end{tabular}
\end{ruledtabular}
\end{table}

\medskip

Existence of the maximal emission angle (\ref{maximal_angle}) leads
to the fact that the differential NCDR cross-section (\ref{dcs-2})
calculated in the laboratory frame becomes infinite when $\theta_f =
\theta_{\rm max}$ [cf. Ref.~\cite{artemyev:03:pra:ncdr} for further
details]. This singularity appears, however, only for the ``model''
calculations with \textit{monochromatic} ions and electrons. If one
considers, in contrast, some velocity distribution of the particles
either in electron target or/and in ion beam such a singularity
transforms into a resonant peak. Theoretical investigation of the
peak structures in the NCDR cross-section is currently underway
and will be reported in a upcoming publication.

\medskip

It immediately follows from Eq.~(\ref{maximal_angle}) and the energy
conservation condition that the maximal positron emission angle
$\theta_{\rm max}$ and, hence, the (position of) singularity in the
recombination cross-section, depends on the energy of the final
helium--like state. This can be observed in the right column of
Fig.~\ref{Fig2} where the NCDR cross-section is displayed in the
laboratory frame as a function of positron emission angle. As seen
from the figure, the positrons emitted in the $K$--shell NCDR have
the largest angle $\theta_{\rm max}$. This angle is drastically
reduced, however, for the electron capture into (weaker bound)
excited ionic states. Such an energy-- and, hence, state--dependence
of the NCDR differential cross-section provides a unique possibility
to separate \textit{experimentally} the positrons which are emitted
in course of the ground-- as well as excited--state electron
recombination. Moreover, the existence of the maximal angle
$\theta_{\rm max}$ may allow one to distinguish NCDR from the
other---competitive---processes which also result in positron
emission. Obviously, this becomes possible if one starts to detect
the outgoing positrons at the polar angles larger then the maximal
one for the NCDR into the ground state.

\medskip

Up to the present we have discussed the angle--differential NCDR
cross-sections both in the projectile (ion--rest) and in the
laboratory (electron--rest) frames. Integration of these cross
sections over the positron emission angle and summation over all
configurations (see
Eqs.~(\ref{cross_section_total})--(\ref{cross_section_partial}))
will yield the \textit{total} NCDR cross-section. In contrast to the
angle--differential ones, this cross-section does not depend on the
particular frame [cf. Ref.~\cite{artemyev:03:pra:ncdr} for further
discussion]. Hence, one can perform the integration over the
positron emission angles in both the ion-- and the electron--rest
frames. Below, for example, the total cross-sections are calculated
within the projectile frame which is more preferable from the
computational viewpoint because of the absence of the singularities
in the positron angular distribution.

\begin{figure}
\vspace*{0.0cm} \epsfig{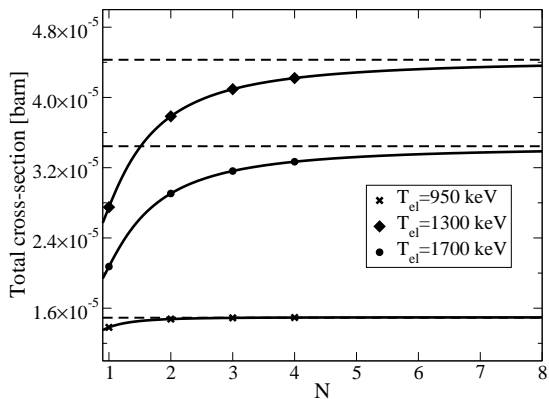}
\caption{\label{Fig3} The convergence of the
NCDR cross-sections $\sigma_N$ (cf. Eq. (\ref{fitting_cross_sections}))
with increasing of $N$ and it's
extrapolation. The symbols are numerical results, the solid lines
represent their interpolation by the third degree polynomials in
inverse powers of $N$, and the dashed lines are the limits of these
polynomials at $N\rightarrow \infty$.}
\end{figure}

\medskip

As seen from Eq.~(\ref{cross_section_total}), the computation of the
total NCDR cross-section $\sigma_{\rm tot}$ requires first a
knowledge of the partial cross-sections $\sigma(n_a \kappa_a, n_b
\kappa_b, J)$ describing the electron recombination into the
particular two--electron states $\ketm{n_a \kappa_a, n_b \kappa_b;
J}$. In our present work, for instance, we take into account all
singly--excited $(1s \, nl_j)$ (up to $(1s \, 4f_{7/2})$) and
doubly--excited $(2s \, 2l_j)$ states of finally helium--like
uranium U$^{90+}$ ions. The partial cross-sections for the
recombination into these states are given in Table \ref{tabU} for
three energies of incoming electron: $T_{kin} = $ 950 keV, 1300 keV
and 1700 keV. As seen from the table, the probability of the NCDR
process significantly reduces for the highly excited ionic states.
For instance, the cross-section of the recombination into the $(1s
\, 4f_{7/2})_{3, 4}$ levels is about four orders of magnitude
smaller comparing with the ground--state one $\sigma((1s)^2)$. Such
a behavior of the NCDR probabilities allowed us to apply an
\textit{extrapolation} scheme in order to estimate the contribution
from the $(1s \, nl_j)$ states with $n>4$ to the total cross-section
(\ref{cross_section_total}). That is, by employing the polynomials
in inverse powers of $N$ for fitting the cross--sections
\begin{equation}
   \label{fitting_cross_sections}
   \sigma_{N} = \sum_{n=1}^N \sum_{l j J} \sigma(1s \, n l_j ; J) \, ,
\end{equation}
we were able to find $\sigma_{\rm tot} = \lim_{N \to \infty}
\sigma_{N}$. Application of the $(1/N)^l$--polynomial fitting seems
to be justified since both the (ordinary) electron capture and
bound--state pair production cross sections are known to scale
roughly as $1/N^3$ \cite{eichler:95:book}. One may expect,
therefore, similar $N$--dependence for the NCDR process which leads
to a production of a singly--excited helium--like ions. The
extrapolation procedure is illustrated in Fig.~\ref{Fig3}, where the
symbols display our numerical results for $\sigma_N$ taken for $N =
1 ... 4$, solid lines plot the interpolation polynomials, and dashed
lines indicate their limits at $N \to \infty$. In contrast to the
singly--excited states, no extrapolation procedure has been applied
for the high--$n$ doubly--excited states since their contribution to
the total NCDR cross-section (\ref{cross_section_total}) was found
to be negligible. Instead, we just added partial cross-sections
$\sigma(2s \, 2l_j)$ to the results of the extrapolation over $(1s
\, nl_j)$ states.

\medskip

In order to estimate an accuracy of our extrapolation scheme, we
have used first only the levels with $n = $ 1, 2 and 3 in order to
find a fitting function. Then we repeated the same procedure but now
including cross sections for the recombination into the $(1s \,
4l_j)$ states. The second asymptotic limit was found to be in about
1--2\% agreement with the first one.

\medskip

The total NCDR cross-sections $\sigma_{\rm tot}$ obtained by means
of the procedure discussed above are given in the last row of Table
\ref{tabU} together with the estimations of the computational error.
These estimations have been derived by comparing the results of the
extrapolation procedure performed with the help of the polynomials
of different degrees. As seen from the last row of the table,
negative continuum dielectronic recombination into excited ionic
states results in a significant enhancement of the total cross-section
$\sigma_{\rm tot}$ when compared with the ground--state capture (cf.
first row of the table). Moreover, the role of the excited states in
the NCDR process increases with increasing kinetic energy
of the electron. While, for example, the excited--state contribution
to the total cross-section (\ref{cross_section_total}) does not
exceed 8 \% for the electron kinetic energy of $T_{\rm el} = $ 950
keV, it increases to almost 75~\% for $T_{\rm el} = $ 1700 keV. Such
a behavior can be easily explained by the fact that the
\textit{threshold} of the NCDR process depends on the total energy
$E_{ab}$ of the (final) two--electron system
(\ref{wave_function_final}) as:
\begin{eqnarray}
   \label{threshold_definition}
   \vare_{ab}^{\rm th} &=& E_{ab} + mc^2 \, ,
\end{eqnarray}
and, hence, increases for excited ionic states. Therefore, since
close to threshold the probability of the NCDR is very low due to
strong exponential damping of the low-energy positron wave function in the
region close to the nucleus, the role of the excited--state recombination
becomes more pronounced only for rather high collision energies.

\medskip

\begin{figure}
\epsfig{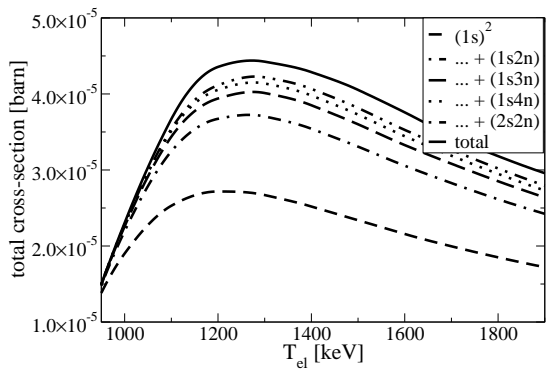}
\caption{\label{Fig4} The total cross-section of NCDR into various
bound states as a function of the kinetic energy of the incoming
electron. }
\end{figure}

In Table~\ref{tabU} we have presented the total
(\ref{cross_section_total}) as well as partial
(\ref{cross_section_partial}) NCDR cross-sections for only three
energies of incoming electron. In order to study the energy
dependence of these cross-sections in more detail we plot them in
Fig.~\ref{Fig4} as a function of $T_{el}$. Similar to before, the
dashed curve in the figure represents the partial cross--section
$\sigma((1s)^2)$ of the NCDR into the ground state. Next four curves
are obtained by adding to $\sigma((1s)^2)$ the capture cross-sections
for the $(1s \, 2l_j)$, $(1s \, 3l_j)$, $(1s \, 4l_j)$, and $(2s \,
2l_j)$ states, respectively [cf. Eq.~\ref{fitting_cross_sections}].
Finally, the solid curve display the total cross-section
$\sigma_{\rm tot}$ obtained as a result of the extrapolation
procedure. Again, the increasing role of the excited ionic state
$(1s \, nl_j)$ and $(2s \, 2l_j)$ for kinetic energies $T_{\rm el}
>$ 1000 keV can be clearly seen from the figure. Apart from the
significant enhancement of the NCDR probability, the recombination
into these states leads also to a slight shift of the maximum of the
total cross-section $\sigma_{\rm tot}(T_{\rm el})$ towards the
higher collision energies.

%
%

%
%
\section{Summary}
\label{sec_summary}

In summary, the negative continuum dielectronic recombination of
high--$Z$, bare ions with free electron targets has been studied
within a framework of the first--order perturbation theory and
Dirac's relativistic equation. For this process, which leads to a
production of a continuum--state positron and a residual
helium--like ion, we have evaluated both, the angle--differential
and the total cross-sections. Special attention in our theoretical
analysis has been paid to the effects on these cross--sections owing
to an electron transition into excited ionic states. We have
demonstrated, for example, that NCDR into singly--excited $(1s \,
nl_j)$ with $n > 1$ and doubly excited $(2s \, 2l_j)$ states of
(finally) helium--like uranium U$^{90+}$ may result in a
significant---up to 75 \%---enhancement of the total recombination
cross-section.

Despite such an enhancement, total recombination cross sections
remain rather small making thus observation of the process difficult
at the present--day facilities. Owing to the recent advances in ion
storage ring and particle detector techniques, however, NCDR
experiments are planed to be carried out at the future ion beam
FAIR facility \cite{GSI01}. One of the signatures of the
NCDR process is that the projectiles change their charge state by
two units. For the estimated NCDR cross sections and beam
intensities of $10^8$ ions/second at an energy of 1.5 GeV/u
(conservative estimate for the FAIR facility \cite{GSI01}), an event
rate of about 2/10 per second can be expected for a target thickness
of $10^{19}$ particle/cm$^2$ . As targets one may prefer low--$Z$
targets such as Be. Here, competitive atomic processes such as
uncorrelated double electron capture might lead to a background of
about one event per minute which may partially mask the original
process. In order to avoid such a background an alternative
experimental set--up may be considered where the emitted positron is
detected in coincidence with the down--charged ions. Assuming a
detector efficiency of the level of 1\% about 10 coincidence events
in one hour could be expected which makes NCDR process clearly
observable at the future FAIR facility in Darmstadt. More detailed
analysis of the various experimental scenarios as well as of the
possible competitive processes is now underway and will be presented
elsewhere \cite{voitkiv}.

%
%
\section{Acknowledgments}

Elucidating discussions with C.~Kozhuharov are gratefully
acknowledged. A.A. and A.S. acknowledge support from the Helmholtz
Gemeinschaft (Nachwuchsgruppe VH--NG--421). The work of V.M.S. was supported
by RFBR (Grant No. 08-02-91967) and by DFG (Grant No. 436RUS113/950/0-1).


\end{document}